\documentclass[twocolumn,preprintnumbers,amsmath,amssymb,superscriptaddress]{revtex4}

\usepackage{graphicx}
\usepackage{dcolumn}
\usepackage{bm}
\usepackage{soul}
\usepackage{color}
\usepackage{epstopdf}
\usepackage[version=3]{mhchem}
\usepackage{lipsum}
\usepackage[outercaption]{sidecap}
\usepackage{floatrow}
\begin{document}
\title{Toward a better understanding of activation volume and dynamic decoupling of glass-forming liquids under compression}
\author{Anh D. Phan}
\email{anh.phanduc@phenikaa-uni.edu.vn}
\affiliation{Faculty of Materials Science and Engineering, Phenikaa University, Hanoi 12116, Vietnam}
\affiliation{ Phenikaa Institute for Advanced Study, Phenikaa University, Hanoi 12116, Vietnam}
\author{Nguyen K. Ngan}
\email{ngan.nguyenkim@phenikaa-uni.edu.vn}
\affiliation{Faculty of Materials Science and Engineering, Phenikaa University, Hanoi 12116, Vietnam}
\author{Nam B. Le}
\affiliation{School of Engineering Physics, Hanoi University of Science and Technology, 1 Dai Co Viet, Hanoi, Vietnam}
\author{Le T. M. Thanh}
\affiliation{Faculty of Basic Science, Posts and Telecommunications Institute of Technology, 122 Hoang Quoc Viet, Hanoi 10000, Vietnam}
\date{\today}

\date{\today}

\begin{abstract}
We theoretically investigate physical properties of the pressure-induced activation volume and dynamic decoupling of ternidazole, glycerol, and probucol by the Elastically Collective Nonlinear Langevin Equation theory. Based on the predicted temperature dependence of activated relaxation under various compression, the activation volume is determined to characterize effects of pressure on molecular dynamics of materials. We find that the decoupling of the structural relaxation time of compressed systems from their bulk uncompressed value is governed by the power-law rule. The decoupling exponent exponentially grows with pressure below 2 GPa. The decoupling exponent and activation volume are intercorrelated and have a connection with the differential activation free energy. We numerically and mathematically analyze relationships among these quantities to explain many results in previous experiments and simulations. 
\end{abstract}

\maketitle
\section{Introduction}
Although the glass transition of amorphous materials has been intensively investigated over three decades, many physical mechanisms and nature underlying this phenomenon remain poorly understood, particularly under effects of pressure \cite{63,64,65,66}. By rapidly cooling materials of melt beyond a threshold rate, the structure of materials falls out of equilibrium and becomes disordered. The non-equilibrium state is characterized by long-range structural disorder. The disordered structures exhibit liquid-like properties at high temperatures but behave as if solid at low temperatures. Obviously, the molecular dynamics or relaxation process is strongly temperature-dependent. The presence of external pressure reduces the free volume known as the volume after excluding the occupied molecular volume and, thus, substantially slows down structural relaxation \cite{63,64,65,66}. In practice, pressure effects occur when manufacturing products in compressors \cite{66,50}. On the one hand, the pressure-induced slowing down can be physically interpreted via changes in differential activation energy and free volume \cite{63,64,9,18}. On the other hand, one can describe the structural relaxation under pressure as a volume-activated process with an activation volume. Gaining insights into a relationship between the activation volume and energy is of importance for industrial and fundamental knowledge. 

Recently, White and Lipson \cite{9} adopted a power law decoupling of the alpha time gradient in confined systems from the bulk time \cite{12,13} to suggest, for the first time, the same behavior in average homogeneous relaxation time of compressed bulk polyvinyl acetate. They used experimental data and simulation of relaxation times to analyze the power law exponent and give simple correlations to the activation energy, free volume, and cooperativity rearrangements. The work also connected confinement effects on glassy dynamics with impacts of pressure on a bulk counterpart. However, there are remaining questions, which are not answered yet, including: (i) does the pressure-induced decoupling apply to other materials? (ii) how does the decoupling exponent depend on external pressure? (iii) how does it relate to the activation volume and the differential energy volume? (iv) can mathematical equations be provided to describe experimental results and simulations?

Among many theories, the Elastically Collective Nonlinear Langevin Equation (ECNLE) theory has provided good quantitative description for the structural relaxation time and dynamic fragility of polymers, thermal liquids, colloidal systems, amorphous drugs, and metallic glasses at ambient and elevated pressures \cite{2,7,10,6,35,42,11,61,62,44,45}. As a minimalist approach, the ECNLE theory employs a hard-sphere fluid to model amorphous materials. Then, the temperature of activated relaxation time ranging from 10 ps to $10^3$s is predicted as a function of particle density. Meanwhile, since the simulation time is limited by $10^6$ ps, simulation cannot quantitatively study experimental observations carried out at 100 s. To contrast ECNLE calculations with experimental data, a density-to-temperature conversion or a thermal mapping is proposed using either experimental equation-of-state data \cite{2,10} or the thermal expansion associated with experimental glass transition temperatures \cite{42,11,35,61,62,44,45}. Clearly, the latter thermal mapping is simpler since only one data is required, but still gives a good agreement of the temperature dependence of activated relaxation and can be applied to various materials.

In this paper, the ECNLE theory is utilized to calculate the temperature dependence of structure relaxation time of ternidazole, glycerol, and probucol at different compression conditions. Then, we use the numerical results to determine the activation volume and differential activation energy. We quantify how molecular dynamics in compressed and uncompressed systems is decoupled via the power law relation and derive an equation of the pressure dependence of the decoupling exponent. Furthermore, this equation is found to capture information of the activation volume and differential activation energy. Correlations among the dynamic quantities are compared to previous simulations and experiments to validate our approach.

\section{Theoretical background}
The ECNLE theory uses a hard-sphere fluid having a volume fraction of $\Phi$ to describe glass-forming liquids \cite{2,7,10,6,35,42,11,61,62,44,45}. When a density of the fluid is sufficiently high ($\Phi \geq 0.432$), a reduction of the free volume decreases the interparticle separation and dynamically localize particles in the fluid. Particles are locally confined within their cages formed by the nearest neighbors. In the absence of external pressure, the dynamic free energy corresponding to the caging localization is
\begin{eqnarray}
\frac{F_{dyn}(r)}{k_BT} &=& -\ln\frac{r}{d}
\label{eq:2}\\ &-&\int_0^{\infty} dq\frac{ q^2d^3 \left[S(q)-1\right]^2}{12\pi\Phi\left[1+S(q)\right]}\exp\left[-\frac{q^2r^2(S(q)+1)}{6S(q)}\right], \nonumber
\end{eqnarray}
where $d$ is a particle diameter in the hard-sphere fluid, $k_B$ is Boltzmann constant, $T$ is temperature, $r$ is the displacement, $q$ is the wavevector, and $S(q)$ is the static structure factor which is computed by the Persus-Yevick theory \cite{1}. 


Equation (\ref{eq:2}) gives us main physical quantities of the local dynamics within a particle cage. The localization of particles is quantified by an emergence of a local barrier emerges in $F_{dyn}(r)$. A minimum and maximum position of $F_{dyn}(r)$ correspond to the localization length, $r_L$, and the barrier position, $r_B$, respectively. The energy difference between these two positions $F_B=F_{dyn}(r_B)-F_{dyn}(r_L)$ is a local barrier.

To diffuse from a cage, surrounding particles are required to rearrange to have an enough space for hopping. The rearrangement causes local vibrations of collective particles quantified by a small displacement field, $u(r)$. This field nucleates from the cage surface and radially spreads over the remaining space. By using Lifshitz's analysis of linear continuum mechanics \cite{5}, we obtain an analytical expression of the displacement field, which is
\begin{eqnarray} 
u(r)=\frac{\Delta r_{eff}r_{cage}^2}{r^2}, \quad {r\geq r_{cage}},
\label{eq:3}
\end{eqnarray}
where $r_{cage}$ is the cage radius determined by the first minimum of the radial distribution function, $g(r)$, and the amplitude of the displacement field, $\Delta r_{eff}$, is \cite{6,7}
\begin{eqnarray} 
\Delta r_{eff} = \frac{3}{r_{cage}^3}\left[\frac{r_{cage}^2\Delta r^2}{32} - \frac{r_{cage}\Delta r^3}{192} + \frac{\Delta r^4}{3072} \right],
\label{eq:4}
\end{eqnarray}
where $\Delta r =r_B-r_L$ is a jump distance.

We suppose that each particle beyond the particle cage harmonically oscillates with an amplitude of $u(r)$ and a spring constant $K_0 = \left|\partial^2 F_{dyn}(r)/\partial r^2\right|_{r=r_L}$. A sum of elastic potential of all oscillators is the elastic barrier quantifying cooperative effects on the glass transition  \cite{2,7,10,6,35,42,11,61,62}. This elastic barrier is calculated by
\begin{eqnarray} 
F_{e} = 4\pi\rho\int_{r_{cage}}^{\infty}dr r^2 g(r)K_0\frac{u^2(r)}{2}, 
\label{eq:5}
\end{eqnarray}
where $\rho$ is the number of particles per volume.

Based on Kramer's theory, the structural relaxation time is
\begin{eqnarray}
\frac{\tau_\alpha}{\tau_s} = 1+ \frac{2\pi}{\sqrt{K_0K_B}}\frac{k_BT}{d^2}\exp\left(\frac{F_B+F_e}{k_BT} \right),
\label{eq:6}
\end{eqnarray}
where $K_B$=$\left|\partial^2 F_{dyn}(r)/\partial r^2\right|_{r=r_B}$ is the absolute curvature at the barrier position and $\tau_s$ is a short time scale \cite{2,7,10,6,35,42,11,61,62}. This short time scale is
\begin{eqnarray}
\tau_s=g^2(d)\tau_E\left[1+\frac{1}{36\pi\Phi}\int_0^{\infty}dq\frac{q^2(S(q)-1)^2}{S(q)+b(q)} \right],
\label{eq:8}
\end{eqnarray}
where $g(d)$ is the contact number, $\tau_E \approx 0.1$ ps is the Enskog time scale \cite{2,6,7,35,42,11,61,62}, $b(q)=\left[1-j_0(q)+2j_2(q)\right]^{-1}$, and $j_n(x)$ is the spherical Bessel function of order $n$.

Using Eqs.(1-6) provides $\tau_\alpha(\Phi)$ while direct comparisons with experiments demand the temperature dependence of structural relaxation time. Thus, it is necessary to have a formula (thermal mapping) to convert from $\Phi$ into $T$ at a certain pressure. In a prior work \cite{44}, based on the thermal expansion process and Hooke's law, a new thermal mapping is formulated as
\begin{eqnarray}
\Phi(P) = \Phi_0e^{-\beta(T-T_0)}\left(1+\frac{P}{E}\right),
\label{eq:9}
\end{eqnarray}
where $P$ is an external pressure, $E$ is an "effective" Young modulus, $\Phi_0 \approx 0.5$ is a characteristic volume fraction, and $\beta \approx 12\times 10^{-4}$ $K^{-1}$ is an effective thermal expansion coefficient for many metallic glasses and organic materials \cite{61,62,11,42,35, 44, 45}. This thermal mapping agrees with
analysis in ref. \cite{40,41}. The value of $\beta$ is deduced from the original thermal mapping \cite{2,10}, which is based on experimental equation-of-state data of many polymers and thermal liquids. In the ENCLE calculations, we find $\tau_\alpha(\Phi_g \approx 0.611) = 100$s and this is connected to the experimental glass transition temperature $T_g$ defined by $\tau_\alpha(T_g)$ = 100s at ambient pressure ($P\approx0$) to determine $T_0$ as \cite{44}
\begin{eqnarray}
T_0 = T_g+\frac{1}{\beta}\ln\left(\frac{\Phi_g}{\Phi_0} \right).
\label{eq:7-1}
\end{eqnarray}
This thermal mapping is theoretically constructed using assumptions of (i) linear elasticity of deformation, (ii) unchanged molecular size, and (iii) a weak pressure dependence of $\beta$. Several experimental literatures studying a triphenylchloromethane/o-terphenyl mixture \cite{30} and poly(vinyl acetate) \cite{31} reveal a small variation of the thermal expansion coefficients under a wide range of applied pressures. Thus, the last assumption seems reasonable but the validity is strongly dependent on material-specific details. 

Overall, there are two characteristic parameters required for our thermal mapping: (i) the experimental $T_g$ at ambient pressure and (ii) the Young's modulus-like value, $E$, to fit with experimental relaxation at higher pressure. All effects of molar mass and particle size on the glass transition are encoded in the $T_g$ value.

\section{Results and discussions}
\begin{figure}[htp]
\includegraphics[width=8.5cm]{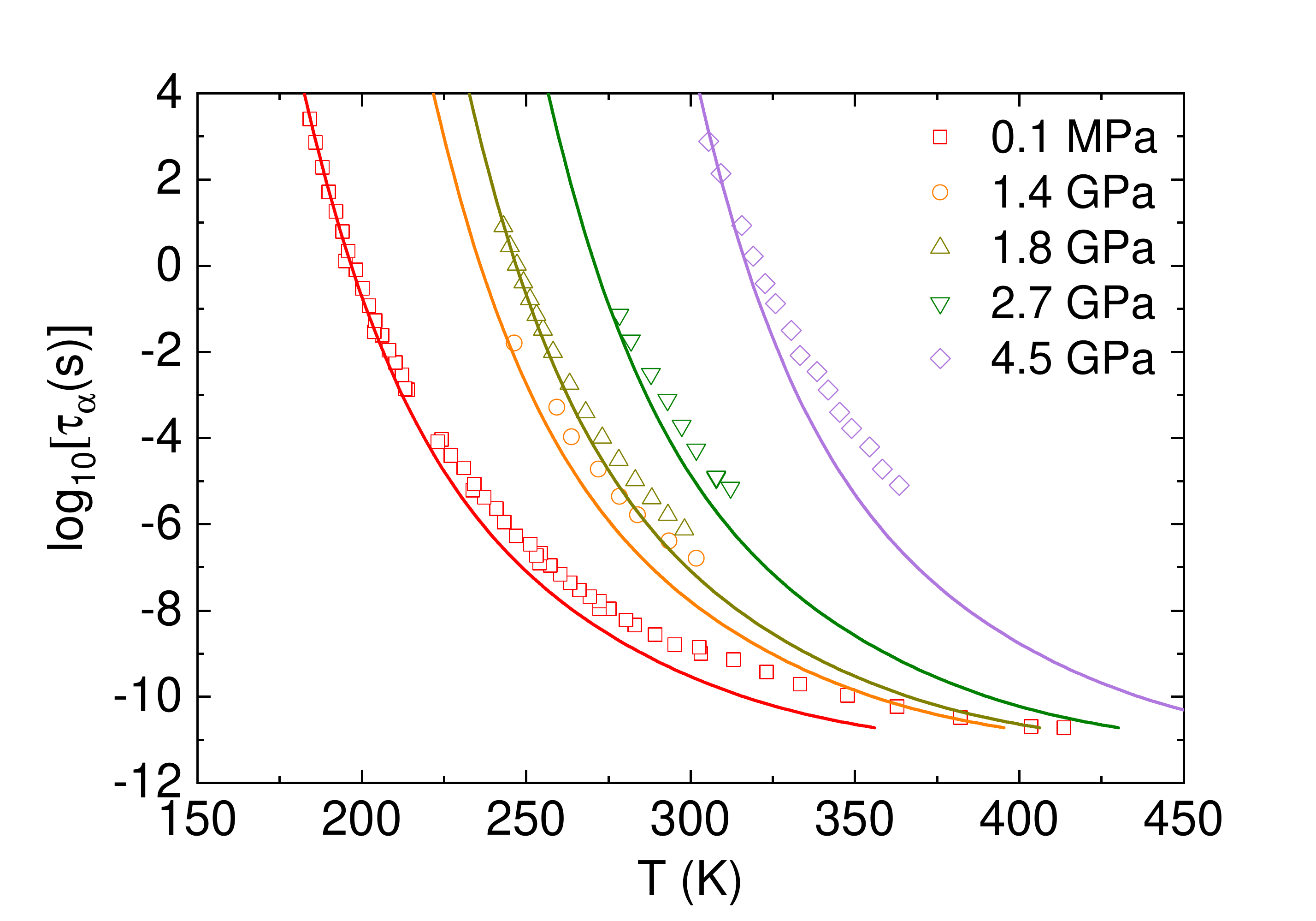}
\caption{\label{fig:0}(Color online) The structural relaxation times of glycerol as a function of temperature at different pressures. Open data points and solid curves corresponds to experimental data in Ref. \cite{24} and ECNLE calculations, respectively.}
\end{figure}

Figure \ref{fig:0} shows the experimental and theoretical temperature dependence of the structural relaxation time of glycerol at $P = 10^{-4}$, 1.4, 1.8, 2.7, and 4.5 GPa. For glycerol, the effective modulus is $E=29$ GPa \cite{44} and $T_g$ at ambient conditions is $190$ $K$. $\tau_\alpha(T,P)$ spanning from 10 ps to $10^4$s are computed by Eqs.(5-8). A pressure-induced decrease of free volume significantly slows down molecular dynamics. The theoretical predictions agree well with experiments for a wide range of pressure and temperature without any adjustable parameter. A deviation theoretical curves and experimental data becomes more remarkable as increasing compression. It suggests that one or some of assumptions of linear elastic deformation, universal correlation between local and collective dynamics, pressure independence of $\beta$, and constant molecular size may be violated at high pressures. Similar agreement and behaviors can be obtained for ternidazole and probucol but the results are not shown here.

The temperature dependence of relaxation time can be used to calculate the dynamic fragility via
\begin{eqnarray}
m = \left. \frac{\partial\log_{10}(\tau_\alpha)}{\partial(T_g/T)}\right |_{T=T_g}.
\label{eq:fragility}
\end{eqnarray} 
Equation (\ref{eq:fragility}) gives us $m = 52.5$ and 63.7 at $P = 0.1$ MPa and 1.8 GPa, respectively. Remarkably, prior experiments \cite{24} found the same values as theory, particularly, $m(P = 0.1 MPa) = 52.2$ and $m(P = 1.8 GPa)=67.6$.

At a given temperature, applying pressure grows both local and elastic barrier and leads to the slowing down of mobility as can be seen from Eqs. (\ref{eq:6}) and (\ref{eq:9}). Empirically, one has another way to measure the sensitivity of the glassy dynamics to external pressures. An increase of the relaxation time can be quantified by the activation volume, $\Delta V^\#$, which is determined by 
\begin{eqnarray}
\tau_\alpha(T,P) = \tau_\alpha(T,P=0)\exp{\left(\frac{P\Delta V^\#}{k_BT} \right)},
\label{eq:10}
\end{eqnarray} 
or
\begin{eqnarray}
\Delta V^\# =\frac{k_BT}{P}\ln\left[\frac{\tau_\alpha(T,P)}{\tau_\alpha(T,P=0)}\right]
\label{eq:10-1}
\end{eqnarray} 

\begin{figure*}[htp]
\center
\includegraphics[width=17cm]{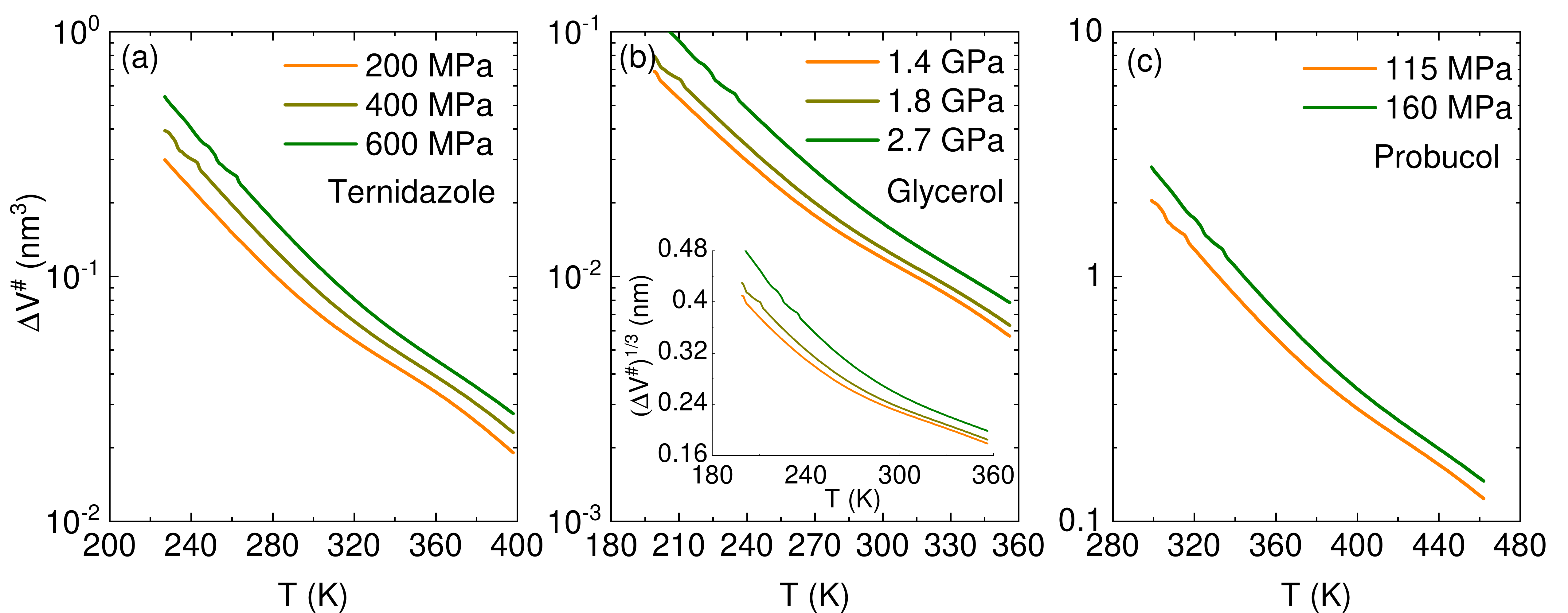}
\caption{\label{fig:1}(Color online) The temperature dependence of activation volume for (a) ternidazole, (b) glycerol, and (c) probucol at elevated pressures. The experimental $T_g(P=0)=$ 294.6 and 231.9 $K$ for probucol and ternidazole, respectively \cite{71,72}. The values of effective Young's modulus $E=$ 1.9 and 8.1 GPa for probucol and ternidazole, respectively \cite{44}. The inset shows our theoretical $(\Delta V^\#)^{1/3}$ of glycerol versus temperature.}
\end{figure*}

Figure \ref{fig:1} shows the temperature dependence of $\Delta V^\#$ of several glass-forming liquids at different pressures. The activation volume monotonically grows with cooling and overall this variation is qualitatively consistent with prior experiments \cite{4,8,17} and simulations \cite{8,3,14}. Our activation volumes are the same order of magnitude as simulations in ref. \cite{3, 14}. On one hand, we roughly observe two linearly decreasing trends of $\log_{10}\Delta V^\#$ occurring in ternidazole, glycerol, and probucol. On the other hand, as seen in the inset of Fig. \ref{fig:1}b, we find that $(\Delta V^\#)^{1/3}$ of glycerol linearly decreases with increasing temperature near $T_g$. This linearity was also reported in ref. \cite{8}.

\begin{figure*}[htp]
\includegraphics[width=17cm]{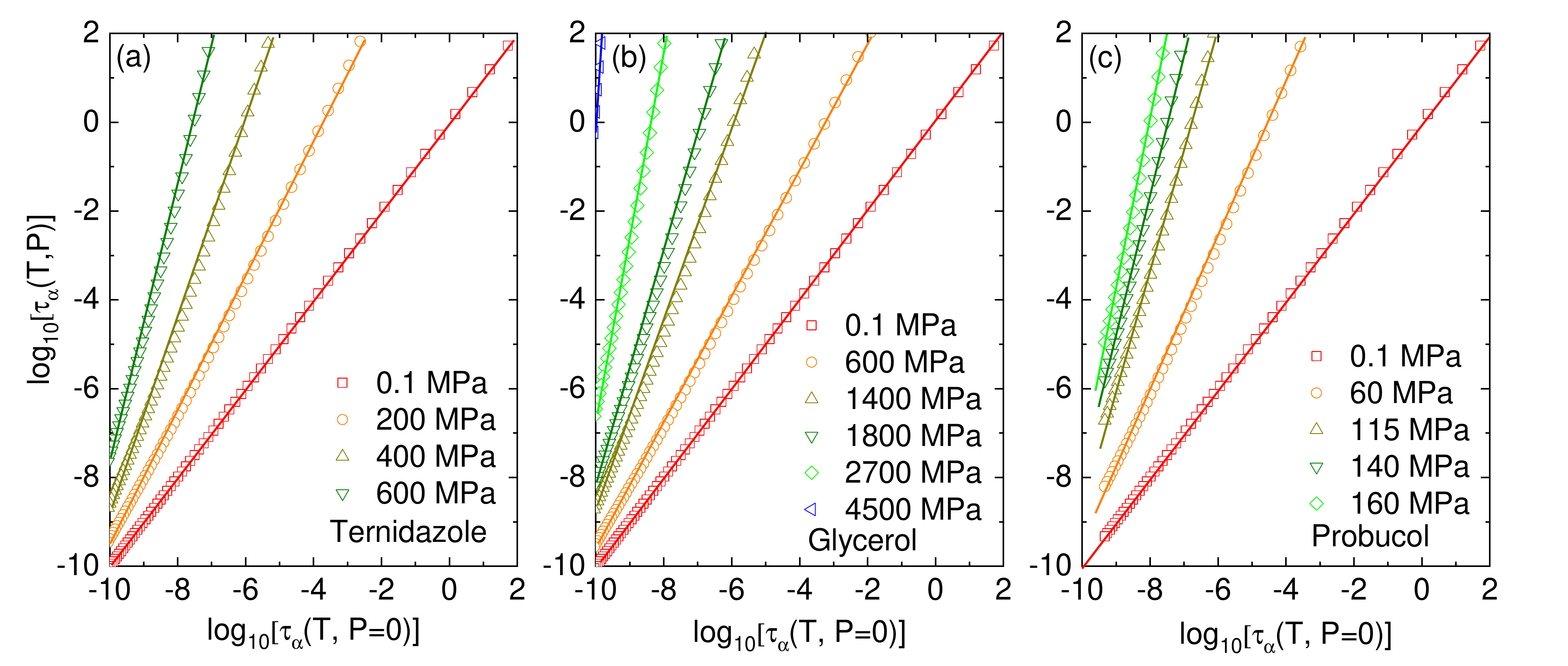}
\caption{\label{fig:2}(Color online) The log of relaxation
times for (a) ternidazole, (b) glycerol, and (c) probucol at elevated pressures as a function of those at atmospheric conditions. Open data points are theoretical calculations and solid lines are guide-to-the-eye lines.}
\end{figure*}

The relaxation time of a compressed system can be correlated to a corresponding bulk as \cite{9}
\begin{eqnarray}
\tau_\alpha(T, P) &=& \tau_\alpha(T,P=0)^{c(P)},
\label{eq:11}
\end{eqnarray}
where $c(P)$ is a pressure-induced decoupling exponent. An idea of the decoupling correlation has been originally used for confined systems \cite{12,13} but, in a recent work, White and Lipson extended this correlation to describe compressed polymer systems \cite{9}.

\begin{figure}[htp]
\includegraphics[width=9cm]{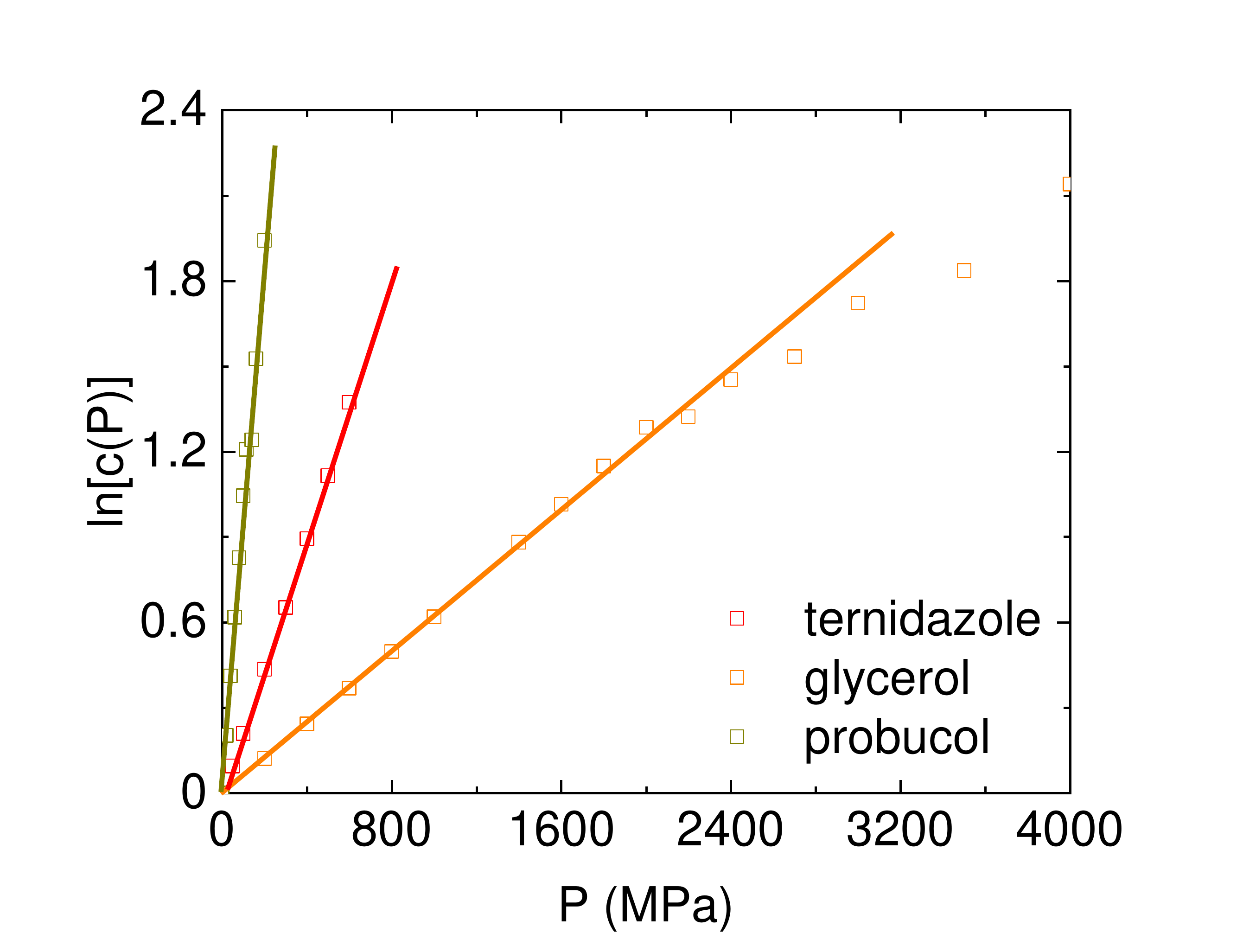}
\caption{\label{fig:3}(Color online) The pressure dependence of the decoupling exponent. Open points and solid lines are our full calculations and guide-to-the-eye lines, respectively.}
\end{figure}

Figure \ref{fig:2} shows a relationship between $\tau_\alpha(T, P)$ and $\tau_\alpha(T, P=0)$ over a wide range of the bulk alpha relaxation time varying by 12 decades. Since almost perfect lines cross through ECNLE data points, there is only a decoupling exponent $c(P)$ at a given pressure and this can be determined by obtaining a slope of the double-log representation. This finding implies simulation can quantitatively predict the decoupling phenomenon in experiments. One can see the pressure dependence of the decoupling exponents of ternidazole, glycerol, and probucol in Fig. \ref{fig:3}. When { $P \leq 2$ GPa, $c(P)$ linearly grows with pressure and the decoupling exponent is approximately expressed by $\ln[c(P)] \approx aP$, here $a$ is a fitting parameter and is insensitive to $P$. A stronger compression leads to a more significant deviation between our ECNLE calculations and the fitting line.}

Equation (\ref{eq:11}) can be transformed to be
\begin{eqnarray}
\frac{\tau_\alpha(T,P)}{\tau_\alpha(T,P=0)} =\tau_\alpha(T,P=0)^{c(P)-1}.
\label{eq:12-1}
\end{eqnarray}
Taking logarithm of both sides of Eq. (\ref{eq:12-1}) gives
\begin{eqnarray}
c(P) =1+\frac{1}{\ln\left[\tau_\alpha(T,P=0) \right]}\ln\left[\frac{\tau_\alpha(T,P)}{\tau_\alpha(T,P=0)}\right].
\label{eq:12}
\end{eqnarray}
As derived in Eq. (\ref{eq:10-1}), $\ln\left[\frac{\tau_\alpha(T,P)}{\tau_\alpha(T,P=0)}\right]$ can be replaced by $\frac{P\Delta V^\#}{k_BT}$. Thus, Eq. (\ref{eq:12}) becomes 
\begin{eqnarray}
c(P) = 1+\frac{P\Delta V^\#}{k_BT\ln\left[\tau_\alpha(T,P=0) \right]}.
\label{eq:13}
\end{eqnarray}
Clearly, Eq. (\ref{eq:13}) suggests a relationship between the activation volume and the decoupling exponent. Since $c(P)\approx e^{aP}$ at pressures less than 2 GPa, the fitting parameter, $a$, is
\begin{eqnarray}
a =\frac{1}{P}\ln\left(1+\frac{P\Delta V^\#}{k_BT\ln\left[\tau_\alpha(T,P=0) \right]}\right).
\label{eq:14}
\end{eqnarray}
When the external pressure is small, we obtain
\begin{eqnarray}
a =\frac{\Delta V^\#/k_BT}{\ln\left[\tau_\alpha(T,P=0) \right]}.
\label{eq:15}
\end{eqnarray}
As seen in Fig. \ref{fig:1}, $\Delta V^\#$ is slightly varied with compression and Eq. (\ref{eq:15}) explicitly explains why $a$ is weakly pressure-dependent. At large pressure, the parameter $a$ depends more on pressure and this explains the presence of non-linearity of $\ln[c(P)]$ with respect to pressure in Fig. \ref{fig:3}.

In a recent work \cite{14}, Xu and his coworkers used simulation to gain insight into a linear relation between $\Delta V^\#(T)$ and
the \emph{differential activation free energy}, $\Delta E_{diff}$, at temperature $T$ given by 
\begin{eqnarray}
\Delta E_{diff} = k_B\frac{\partial\ln(\tau_\alpha)}{\partial(1/T)}.
\label{eq:16}
\end{eqnarray} 
Comparing Eq. (\ref{eq:fragility}) with Eq. (\ref{eq:16}), the differential activation free energy is mathematically related to the dynamic fragility via $m=\Delta E_{diff}/(k_BT_g\ln10)$ and determines the slope in an Arrhenius plot of $\tau_\alpha$ at $T_g$. Equation (\ref{eq:16}) leads to an approximate expression of the relaxation time $\tau_\alpha = \tau_0e^{\Delta E_{diff}/k_BT}$ with $\tau_0$ being the time at infinite temperature. After combining this approximation with Eq.(\ref{eq:15}), we have
\begin{eqnarray}
\Delta V^\# &=& ak_BT\ln\left[\tau_\alpha(T,P=0) \right] \nonumber\\
&=& a\left(k_BT\ln\tau_0 + \Delta E_{diff} \right).
\label{eq:17}
\end{eqnarray}
Equation (\ref{eq:17}) explicitly indicates that $\Delta V^\#$ is linearly proportional to $\Delta E_{diff}$. Our calculations provide better physical analysis than simulation in ref. \cite{14} because of straightforward mathematical transformations. Moreover, this finding is consistent with previous experimental results \cite{15,16}.

\begin{figure*}[htp]
\center
\includegraphics[width=17cm]{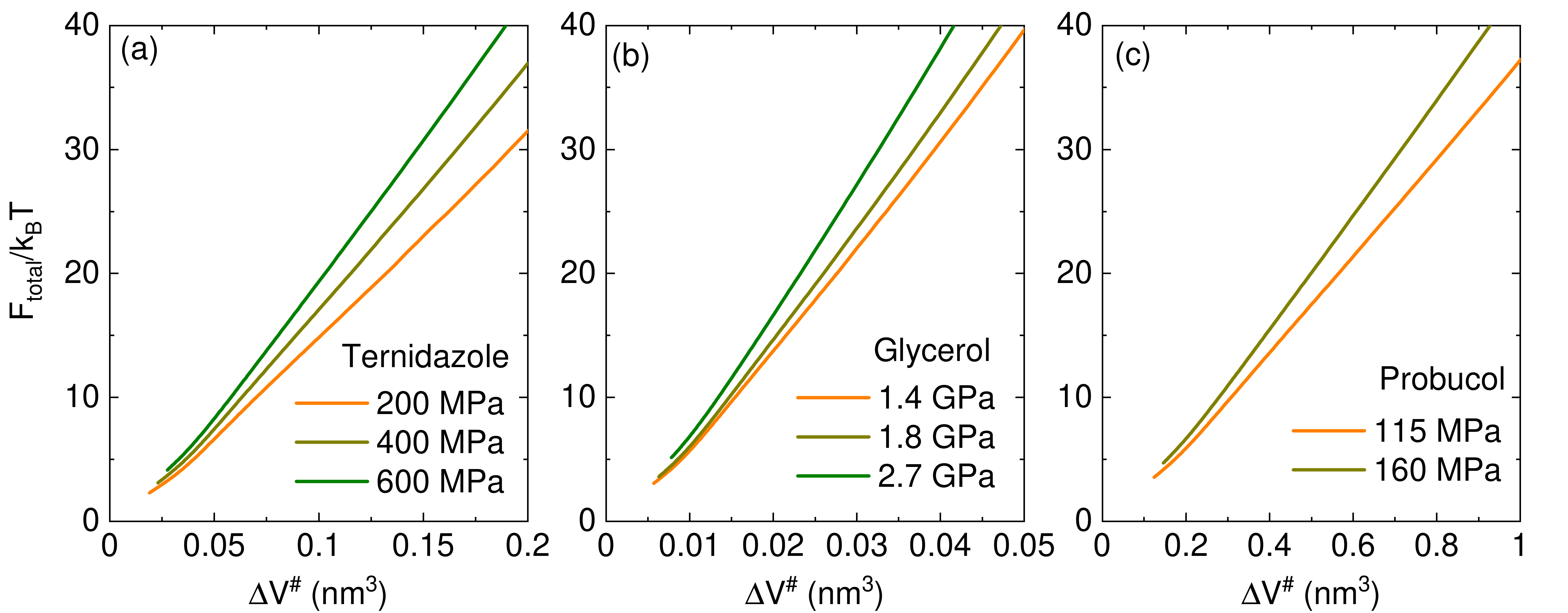}
\caption{\label{fig:4}(Color online) The total barrier normalized by $k_BT$ as a function of the activation volume of (a) ternidazole, (b) glycerol, and (c) probucol at elevated pressures.}
\end{figure*}

In the framework of the ECNLE theory, we can consider $\Delta E_{diff}$ as the total barrier $F_{total}=F_B+ F_e$. The proportionality between $E_{diff}$ and $\Delta V^\#$ can be quantified via $F_{total}/k_BT$ as a function of $\Delta V^\#$. Thus, we calculate these quantities and show numerical results in Fig. \ref{fig:4}. Remarkably, we predict perfect linear increases of the normalized total barrier with increasing the activation volume over a wide range of external pressures. Note that the timescale for determining the activation volume ranges from simulation to experimental scale. These straight lines in Fig. \textcolor{red}{\ref{fig:2}} mean simulation results can be exploited to quantitatively study physical behaviors in experiments.

In recent works \cite{9,18}, White and Lipson connected the dynamics to the free volume, $V_{f}(T,P)= V(T,P) - V_{hc}$, via $\ln\tau_\alpha  \varpropto V_{hc}/V_{f}$, where $V(T,P)$ and $V_{hc}$ are the total volume of systems and the close packing volume, respectively. From this, the decoupling exponent is calculated by \cite{9}
\begin{eqnarray}
c(P) = \frac{\ln\tau_\alpha(T,P)}{\ln\tau_\alpha(T,P=0)} = \frac{V_f(T,P=0)}{V_f(T,P)}.
\label{eq:18}
\end{eqnarray}
To obtain this result, these authors assumed that $V_{hc}$ weakly depends on pressure. This is consistent with our assumption of no change in size and chemical/biological structure of molecular particles under pressure. By the definition of $\Delta E_{diff}$, we have $\ln\tau_\alpha\varpropto\Delta E_{diff}\varpropto1/V_f(T,P)$. Meanwhile, at small pressures, the relationship between the activation and free volume is given by
\begin{eqnarray}
\Delta V^\# &=& ak_BT\ln\left[\tau_\alpha(T,P=0) \right] \nonumber\\
&\varpropto&  ak_BT\frac{V_{hc}}{V_{f}(T,P=0)}\varpropto \frac{1}{V_f(T,P=0)}.
\label{eq:19}
\end{eqnarray}
In addition, since $c(P)=e^{aP}$, the free volume at an elevated pressure can be linked with that at atmospheric pressure by $V_f(T,P) = V_f(T,P=0)e^{-aP}$ or $\ln\left(\frac{V_f(T,P=0)}{V_f(T,P)}\right)\varpropto P$.


\section{Conclusions}
In conclusion, the activation volume and dynamic decoupling exponent are key parameters governing the glassy dynamics under compression. We have used the ECNLE theory calculate the temperature dependence of the structural relaxation time of probucol, glycerol, and temidazole at several pressures. Then, the activation volume is calculated and decreases with heating. The structural relaxation time of compressed systems is deviated from that of uncompressed counterparts in a power law manner, $\tau_\alpha(T, P) = \tau_\alpha(T,P=0)^{c(P)}$ over 12 decades. We predicted that the decoupling exponent varies exponentially with pressure below 2 GPa. At stronger pressure, the elastic behavior may become nonlinear or the molecule size may be changed. After straightforward mathematical transformations, a clear link among $\Delta V^\#$, $c(P)$, and $\Delta E_{diff}$ is established, particularly $\Delta V^\# \sim \Delta E_{diff}$. Our numerical results and analysis have provided qualitative descriptions for many prior simulations and experiments. These findings provide a better understanding of compression effects on the glass transition of glass-forming liquids.

\begin{acknowledgments}
This research was funded by the Vietnam National Foundation for Science and Technology Development (NAFOSTED) under grant number 103.01-2019.318. 
\end{acknowledgments}

\end{document}